\documentclass[12pt,preprint]{aastex}

\slugcomment{To appear in the ApJ Supplements, version 3.0 feb/2006}

\shorttitle{A 2MASS-Cleaned FHB/A Star Catalog}
\shortauthors{Beers et al.}

\begin{document}


\title{A Catalog of Candidate Field Horizontal-Branch and A-Type Stars. III.
 A 2MASS-Cleaned Version}

\author{Timothy C. Beers\altaffilmark{1,2}}
\affil{Department of Physics \& Astronomy, CSCE: Center for the Study of Cosmic Evolution,
and JINA: Joint Institute for Nuclear Astrophysics, Michigan State University, East Lansing, MI 48824, USA}
\email{beers@pa.msu.edu}

\author{Tiago Almeida, Silvia Rossi \altaffilmark{1}}
\affil{Instituto de Astronomia,  Geof\'{i}sica e Ci\^{e}ncias Atmosf\'{e}ricas, Departamento de Astronomia,  Universidade de S\~{a}o Paulo, \\ 
Rua do Mat\~{a}o  1226, 05508-900 S\~{a}o Paulo, Brazil}
\email{almeida@astro.iag.usp.br, rossi@astro.iag.usp.br}

\author{Ronald Wilhelm}
\affil{Department of Physics, Campus Box 41051, Texas Tech University, Lubbock, TX 79409, USA}
\email{ron.wilhelm@ttu.edu}

\and

\author{Brian Marsteller}
\affil{Dept. of Physics \& Astronomy, CSCE: Center for the Study of Cosmic Evolution,
Michigan State University, E. Lansing, MI  48824, USA}
\email{marsteller@pa.msu.edu}

\altaffiltext{1}{Visiting Astronomer, Cerro Tololo Inter-American Observatory.
CTIO is operated by AURA, Inc.\ under contract to the National Science
Foundation.}
\altaffiltext{2}{Visiting Astronomer, Kitt Peak National Observatory.
KPNO is operated by AURA, Inc.\ under contract to the National Science
Foundation.}

\begin{abstract}

We present coordinates and available photometric information (either from
previous or recent broadband $UBV$ observations, and near-infrared photometry
from the 2MASS Point Source Catalog) for 12056 stars (11516 of which are unique)
identified in the HK Survey of Beers and colleagues as candidate field
horizontal-branch or A-type stars. These stars, in the apparent magnitude range
10 $\le B \le 16.0$, were selected using an objective-prism/interference-filter
survey technique. The availability of 2MASS information permits assembly of a
cleaned version of this catalog, comprising likely blue horizontal-branch (BHB)
stars or blue stragglers in the color interval $-0.2 \le (B-V)_0 \le
+0.2$, which are of particular interest for investigation of the structure,
kinematics, and dynamics of the thick disk and inner halo of the Milky Way, the
total mass and mass profile of the Galaxy, and as potential
foreground/background objects in efforts to bracket distances to high
velocity clouds of H~I. A comparison of the stars classified as high-likelihood
BHB candidates with previous classifications based on $UBV$ photometry and
medium-resolution spectroscopy indicates that this class contains 78\% correct
identifications.     

\end{abstract}




\section{Introduction}

It has long been recognized that stars in the horizontal-branch stage of
evolution are attractive probes of the Galactic thick-disk and halo populations,
owing to their relatively large numbers as compared to other blue objects, and
to the ease with which reasonably accurate distances can be determined. 

Until quite recently, the primary discovery method for field horizontal-branch
(FHB) stars has been inspection of wide-field objective-prism plates
obtained during the course of surveys of the Galactic halo (e.g., Pier 1982;
Beers, Preston, \& Shectman 1988, FHB~I; MacConnell, Stephenson, \&
Pesch 1993; Rodgers, Roberts, \& Walker 1993; Beers et al. 1996, FHB~II). The
most recent survey of this type, the Hamburg/ESO Survey (HES) of Christlieb and
collaborators (Christlieb 2003), also identifies large numbers of FHB candidates
(Christlieb et al. 2005). A number of important studies of the structure and
kinematics of the Galaxy have been carried out with previously identified
samples of horizontal-branch stars from prism-survey studies. These include the
work of Pier (1984), Doinidis \& Beers (1989), Kinman, Suntzeff,
\& Kraft (1994), Preston, Shectman \& Beers (1991), Sommer-Larsen et al. (1997),
and Thom et al. (2005).

Photometric selection techniques (e.g., Norris \& Hawkins 1991; Arnold \&
Gilmore 1992; Flynn et al. 1995) have also been used, in particular for the
identification of more distant FHB stars. Such colorimetric selections have been
limited, however, to inspection of small patches of sky. The advent of several
large-scale photometric surveys in the optical (the Sloan Digital Sky Survey --
SDSS, York et al. 2000; Adelman-McCarthy et al. 2006), in the near-IR (2MASS;
Skrutskie et al. 2006), and soon, in the near- and far-UV (GALEX; Martin et al.
2005) are rapidly opening windows on large fractions of the sky for
identification of likely FHB stars. Even before the formal completion of the
SDSS in July 2005, several studies, e.g., Yanny et al. (2000), Newberg et al.
(2003), Sakamoto, Chiba, \& Beers (2003), and Sirko et al. (2004 a,b), made use
of existing SDSS-selected FHB candidates for studies of the kinematics and
dynamics of the inner and outer halo, and to identify the presence of
substructure in the Galactic halo. Clewley et al. (2005) report on
observations of SDSS-selected A-type stars that appear to reveal the present of
a distant Galactic stream some 70 kpc away. Brown et al. (2004) have shown how
photometric selection of FHB candidates from 2MASS photometry can be used for similar
applications. 

For efficiency of subsequent follow-up observations of FHB stars (and for
consideration of their clustering properties), it is particularly important to
obtain as pure a catalog of FHB stars as possible. Given the apparently large
fraction of high-gravity A-type stars in the thick disk and the halo (the great
majority of which are likely to be blue stragglers) that might confound samples
of FHB candidates\footnote {Norris \& Hawkins (1991) estimate that on the order
of 50\% of the blue stars they considered have higher gravities than would be
associated with FHB stars; Wilhelm et al. (1999b) comes to a similar
conclusion.}, one would ideally attempt to implement some gravity-sensitive
probe. Indeed, Christlieb et al. (2005) have reported on just such an effort,
making use of Str\"omgren indices obtained directly from the
wavelength-calibrated HES plates; these authors report that they have achieved a
contamination level of less than 16\% from higher-gravity A-type stars.

The HK survey plates used in the catalogs described in FHB~I and FHB~II were not
wavelength calibrated, nor did they cover as wide a wavelength range as the HES
plates, so unfortunately no gravity separation can be effectively applied.
However, the HK survey FHB/A candidates remain valuable for several reasons: (a)
they extend to brighter apparent magnitudes (hence nearer distances) than
alternative surveys (such as the SDSS, which has a bright limit on the order of
$g \simeq 14$), (b) and they explore directions, in particular in the northern
hemisphere, which are not covered by the HES candidates, and in the southern
hemisphere, which are not covered by the SDSS candidates. 

Over the course of the past few years, additional photometric information (in
particular from the 2MASS survey) has come available for a large number of 
HK-survey FHB/A candidates. In the present paper we make use of this information
to produce a cleaned catalog of HK survey blue horizontal-branch (BHB)
candidates, by identification of the subset of stars with estimated $B-V$ colors
in the range $-0.2 \le (B-V)_0 \le 0.2$. For completeness, we also make use of
available measured $UBV$ photometric information gathered for HK-survey
candidate FHB/A stars over the past two decades. This information, taken as a
whole, allows one to sub-classify the original HK-survey FHB/A candidates into
three primary categories: high-likelihood BHB candidates, medium-likelihood BHB
candidates, and low-likelihood BHB candidates. This classification should prove
useful for future observations of the sub-samples of greatest interest from the
HK survey.

In \S 2 we describe the selection of the FHB/A candidates and the assembly of
our catalog. Section 3 describes the methods we use to obtain estimated $V$
magnitudes and $B-V$ colors, based on 2MASS photometry, for catalog stars that
lack this information. In \S 4 we discuss the likelihood assignments for the HK
survey BHB candidates. The catalog of BHB candidates is presented in \S 5. A
brief discussion, and plans for future observations of catalog stars, is given
in \S 6.

\section{Candidate Selection and Catalog Assembly}

Details of the objective-prism/interference-filter technique employed in the HK
survey have been provided elsewhere (see FHB~I, FHB~II, and references therein),
and will not be repeated here. FHB~II presents a discussion of the small
differences in the survey techniques for the northern- and southern-hemisphere
samples arising from the different emulsions and interference filters employed.
We do not explicitly consider these differences in the present catalog.

A total of 132 acceptable plates located in the northern Galactic hemisphere and
169 plates located in the southern Galactic hemisphere were obtained during the
course of the HK survey, 278 of which are unique (that is, without substantial
overlap with other survey plates), for a total sky coverage of roughly 7000
square degrees. Table 1 lists the full objective-prism plate log from which
stars that have been incorporated into the present catalog are drawn. This Table
supercedes the logs previously presented in FHB~I and FHB~II. Note that our
present catalog includes some 3300 FHB/A candidates that were not included in
either of these previous catalogs. Table 1 lists the plate numbers, equatorial
coordinates, Galactic coordinates of the plate centers, and the total number of
FHB/A candidates identified on each plate (including all of the stars originally
classified as FHB/A, not just those within the restricted color ranges discussed
below). The plate centers and Galactic coordinates of the plate centers were
obtained by averaging the positions of the FHB/A candidates contained on them.
The final column of Table 1 indicates the plates that have centers within $2^o$
of one another, and thus have substantial overlap. Figure 1 shows the positions
of all of the HK-survey plates used in the assembly of the present catalog.

\begin{figure}
\epsscale{0.75}
\plottwo{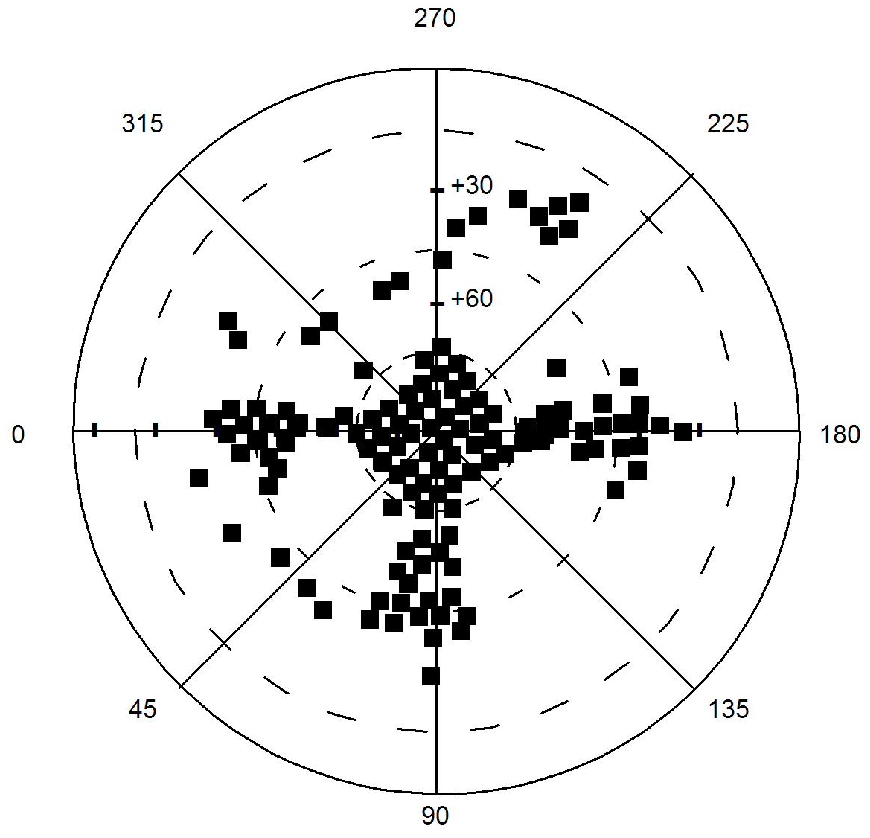}{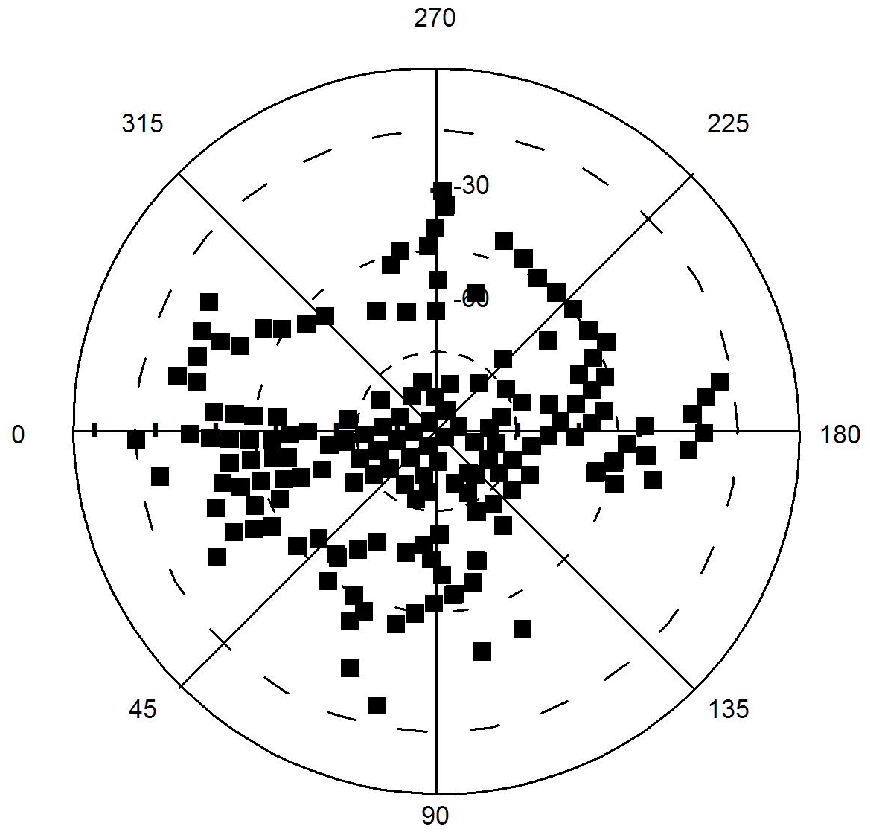}
\caption{(Left Panel) Location of the HK-survey objective-prism plates on a 
polar plot of the northern Galactic hemisphere. The plot is centered on the
solar position; $l=0^o, b=0^o$ is the direction of the Galactic center; $l=90^o,
b=0^o$ is the direction of Galactic rotation. The boxes are drawn roughly to
scale. (Right Panel) A similar plot indicating the positions of HK-survey plates
in the southern Galactic hemisphere.}
\label{fig1}
\end{figure}

During the course of the visual selection and classification of candidates, the
classifiers (Preston and Beers) assigned broad types within which the majority
of FHB/A stars are likely to fall. These types were assigned based on the
strength of the Balmer H-$\epsilon$ line and the CaII K~line, both of which fell
within the narrow 150 \AA\ window of spectrum admitted by the HK interference
filter. These classes were AB, indicating a broad H-$\epsilon$ feature and
absence of the CaII~K line, and A, indicating a broad H-$\epsilon$ feature and
the presence of a weak- to moderate-strength CaII~K line. Such distinctions become more
difficult to make as the candidates approach either the bright limit ($B \sim
10.0$, due to saturation of the photographic emulsion) or the faint limit ($B
\sim 16.0$, due to rising plate noise) of the survey. The classifiers also
assigned discrete brightness estimates, as described in FHB~I and FHB~II. While
useful for initial inspection of the catalogs, these brightness estimates
clearly are subject to potentially large plate-to-plate variations, and to the
personal biases of the classifiers.

Fortunately, we are now in a position to make use of a substantial amount of
new photometric information, either from $UBV$ observations of catalog objects
that has been conducted over the course of the past few decades (including
recent observations; see Beers et al. 2006) or, in particular, near-IR
photometry from the 2MASS Point Source Catalog.

\subsection{Matches with 2MASS}\label{match2mass}

The coordinates of candidates measured from the HK-survey plates have an
estimated r.m.s. scatter on the order of 2"-3", with occasional much larger
deviations (up to 15") for stars near the edges of the prism plates. Hence, in
order to search the 2MASS catalog, we initially obtained matches using a search
radius of 20". This resulted in the identification of $\sim 99$\% of the stars
in our complete catalog listing; roughly 20\% of these matches resulted
in more than one star from the 2MASS catalog being selected as a possible match.
We consider a (single) star to be a successful match if the difference in the
input (survey) coordinates and the 2MASS coordinate is less than 10", and the
reported 2MASS $J$ magnitude is roughly commensurate with the crude brightness
estimates obtained during the visual classification process (see FHB~II for
further discussion of the brightness classifications).

\begin{figure}
\epsscale{0.9}
\plotone{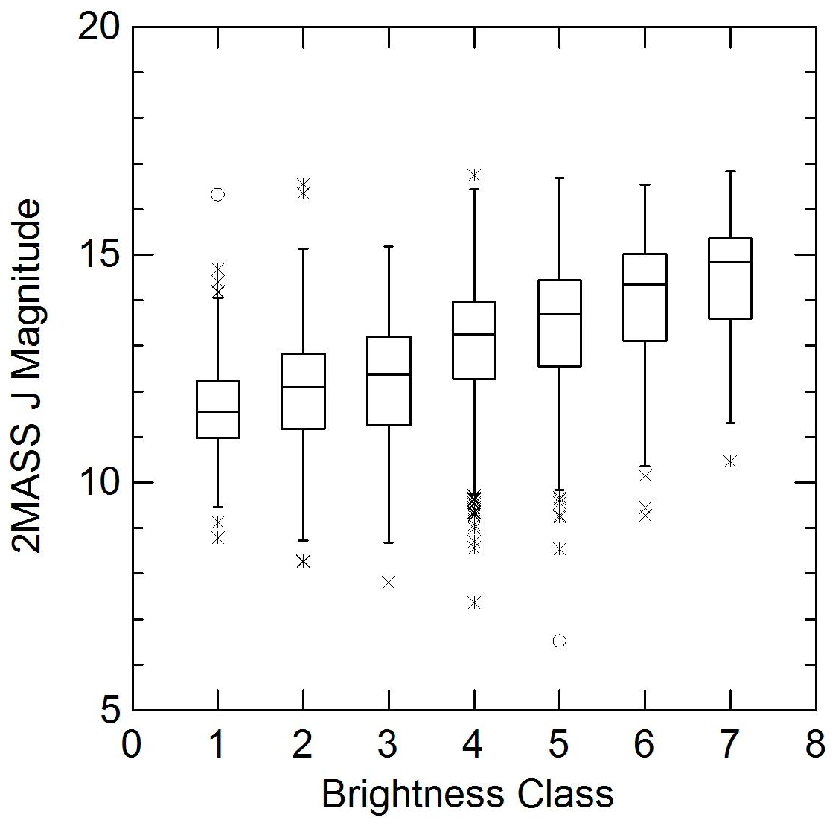}
\caption{Boxplots illustrating the relationship
between the 2MASS $J$ magnitude and assigned brightness classes.
Here we use a numerical brightness classification scale, from 1 to 7. 
This corresponds to the brightness classes discussed in FHB~I and FHB~II as
follows:  (1) vb, (2) b, (3) mb, (4) m, (5) mf, (6) f , (7) vf.  For each
brightness class, the box indicates the interquartile range, within which 50\%
of the data lie.  The horizontal line in the middle of each box is placed at the
location of the median apparent magnitude.  The vertical lines on either side of
the box extend to include the last data points not considered outliers. Outliers
are illustrated with asterisks (indicating a mild outlier) or open circles
(indicating a severe outlier).}
\label{fig2}
\end{figure}

Figure 2 shows boxplots of the correlation between the assigned brightness
classes for the complete set of initial (single star) matches with 2MASS $J$
magnitude. As can be seen from inspection of this Figure, the brightness
classifications roughly predict the corresponding 2MASS $J$ magnitudes, albeit
with a rather large scatter. Note that some of this scatter is attributable to
the fact that the candidates cover a rather wide range in color. Table 2
summarizes robust estimates of the central location ($C_{BI}$,``mean'') and
scale ($S_{BI}$, ``dispersion'') in the distributions of 2MASS $J$ magnitudes
associated with each brightness class (where the ``BI'' subscript indicates the
use of the biweight estimators described by Beers, Flynn, \& Gebhardt 1990). As
discussed in FHB~II, we expect that there exists small differences in the
typical $J$ magnitudes between stars obtained from plates where different
interference filters and photographic emulsions were employed (as well as from
the different personal biases of the classifiers). For the purpose of the
matching exercise we choose to ignore these small shifts, but they will
obviously add some noise to the procedure. We consider a star to have a
commensurate 2MASS $J$ magnitude if the the difference between the $J$ magnitude
associated with its brightness class and the reported $J$ magnitude is less than
1.5 magnitudes. This criterion eliminates no more than 3\% of the initial
matches.

All multiple identifications of FHB/A candidates with the 2MASS catalog were
inspected in more detail. In the vast majority of cases, the closest star among
the possible matches is clearly the correct choice, based on the comparison of
expected $J$ magnitude described above. In rare cases more than one match is
possible, since more than a single star passes this comparison, and a confident
choice is not possible.  A number of these multiple matches are expected, based
on the overlap of prism spectra of similar magnitudes on the original survey plates,
in particular when the stars are approximated aligned along a north/south
axis, which is in the dispersion direction of the HK-survey plates. Such
stars are noted in the catalog listing described in \S 5 below.

\section {Estimation of $V$ Magnitudes and $B-V$ Colors}

We now make use of the subset of successfully matched stars from the procedure
described above and with available broadband $UBV$ photometry (from Doinidis \&
Beers 1990, 1991; Preston, Shectman, \& Beers 1991; Norris, Ryan, \& Beers 1999;
Bonifacio, Monai, \& Beers 2000; Beers et al. 2006) to obtain
predictions of $V$ magnitudes and $B-V$ colors based on 2MASS $J$ and $J-H$
colors. First, we apply reddening corrections taken from Schlegel, Finkbeiner,
\& Davis (1998).  Note that for some stars at low Galactic latitudes, the 
Schlegel et al. reddenings ($E(B-V)_S \ge 0.10$) have been revised slightly
downward following the prescription of Bonafacio et al. (2000):

$$E(B-V)_A = 0.10 + 0.65 * (E(B-V)_S - 0.10)$$

Here the subscript ``A'' indicates the adopted reddening. Figure 3a shows
residuals obtained from the correlation between measured $V_0$ magnitudes
(corrected for extinction assuming A(V) $= 3.1 E(B-V)_A$) and predicted 2MASS
$V_{0,2M}$ magnitudes. This regression was obtained for the 606 stars with
available $V$ magnitudes and located in regions with adopted reddening estimates
$E(B-V)_A \le 0.03$. A robust (least median of squares) regression relation of
the form:

$$V_{0,2M} = a_0 + a_1*J_0 + a_2*(J-H)_0 + a_3*(J-H)_0^2$$

\noindent with coefficients:  a$_0$ = 0.599 (0.050), a$_1$ = 0.966 (0.030),
a$_2$ = 2.645 (0.100), a$_3$ = 0.927 (0.200) has $R^2$ = 0.98 and $\epsilon $=
0.14, where $R^2$ indicates the fraction of the variance in the estimate that is
accounted for by the correlation, and $\epsilon$ indicates the expected error of
the prediction (in magnitudes). The parenthetic quantities following each
of the coefficients are the one-sigma estimates of their errors. Note that the use
of a robust regression approach means that this relationship is not unduly
affected by the presence of outliers.

Figure 3b shows residuals obtained from the correlation between measured $(B-V)
_0$ and the predicted 2MASS $(B-V)_{0,2M}$ color, for the 606 stars with
available (B-V) colors and located in regions of
low reddening. A robust regression relation of the form:

$$(B-V)_{0,2M} = a_0 + a_1*(J-H)_0 + a_2*(J-H)_0^2$$

\noindent with coefficients:  a$_0$ = 0.021 (0.002), a$_1$ = 0.913 (0.023),
a$_2$ = 2.129 (0.21) has $R^2$ = 0.74 and $\epsilon$ = 0.08. As before, the
parenthetic quantities following each of the coefficients are the one-sigma
estimates of their errors. As indicated by the small value of $\epsilon$,
the above relationship provides a useful means of obtaining an estimated $(B-V)
_0$ color for the FHB/A candidates from which we wish to select likely BHB
candidates. Note that the typical errors on the known $(B-V) _0$ colors ($\sim
0.01-0.02$ mags) and the 2MASS $(J-H)_0$ colors ($\sim 0.04$ mags) can account
for roughly half of this scatter. Errors in the reddening estimates add another
source of scatter.

\begin{figure}
\epsscale{0.75}
\plottwo{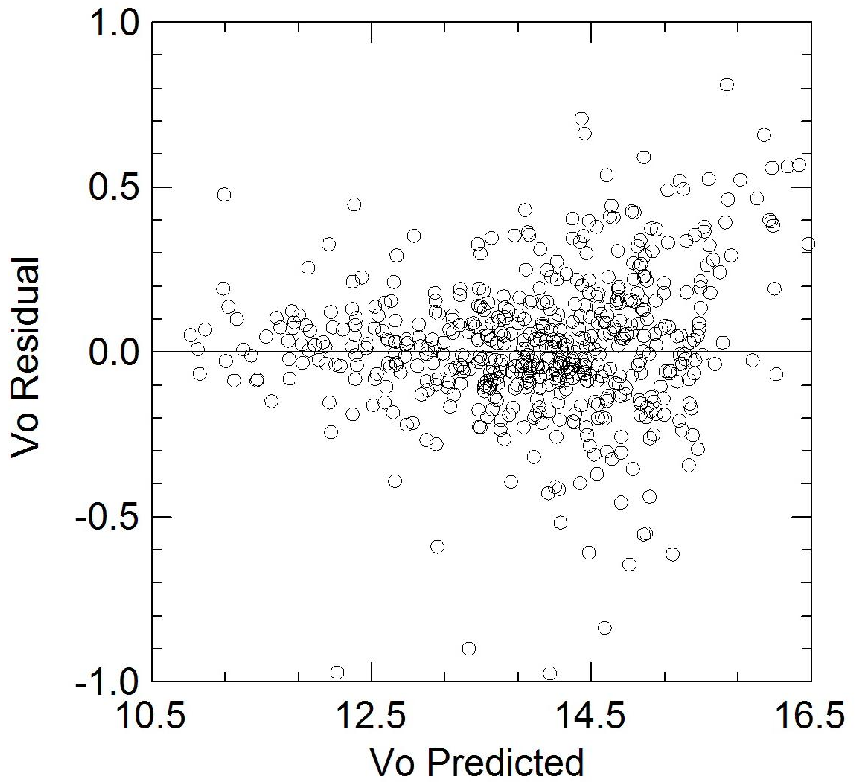}{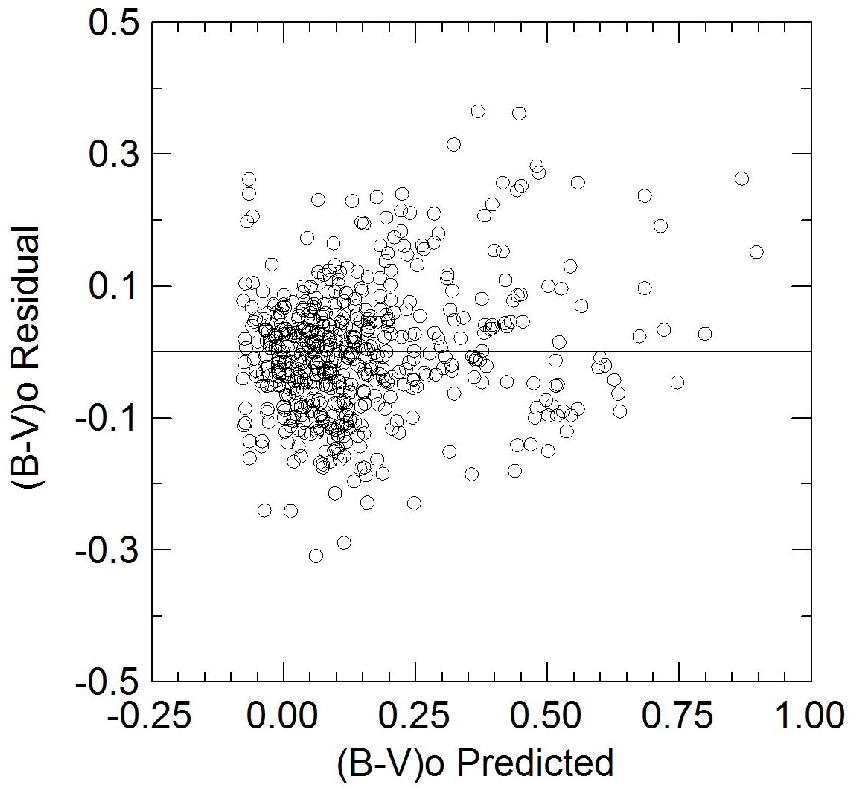}
\caption{(Left Panel) Residuals in the estimate of $V_0$ based on 2MASS
photometry as a function of the predicted $V_0$.  (Right Panel) 
Residuals in the estimate of $(B-V)_0$ based on 2MASS
photometry as a function of the predicted $(B-V)_0$.}
\label{fig3}
\end{figure}

\section {Assignment of BHB Likelihood}

The classifications of candidate FHB/A stars in the HK survey extend over a
rather broad color range, $-0.3 \le B-V \le 0.6$. Although there surely exist
{\it bona-fide} FHB stars over this entire range, in the redder portion there
also exists a substantial number of stars that are not BHB stars,
but rather, are metal-deficient main-sequence or subgiant stars located near the
turnoff of the halo- or thick-disk populations that were mistaken for hotter
stars in the initial prism-spectra classifications. At the bluer end of this range,
there also exist sdB and halo blue stragglers that can be confused
with BHB stars (see the discussion of these points in Wilhelm et al. 1999a).
Hence, we choose to isolate the subset of our catalog objects that have the
greatest likelihood of being {\it bona-fide} BHB stars. This assignment is of
course only approximate, but it should prove useful for prioritizing the targets
of greatest interest for future spectroscopic and photometric follow-up
observations.

The BHB likelihood is assigned in the following manner:

\begin{itemize}

\item High (H):  The star has either a measured or estimated $(B-V)_0$ color in the range
$-0.2 \le (B-V)_0 \le +0.2$

\item Medium (M):  The star has either a measured or estimated $(B-V)_0$ color in the range
$-0.3 \le (B-V)_0 < -0.2$ or $+0.2 < (B-V)_0 \le +0.4$

\item Low (L):  The star has either a measured or estimated $(B-V)_0$ color in the range
lying outside of $-0.3 \le (B-V)_0 \le +0.4$

\end{itemize}

Given the errors in the 2MASS $(J-H)_0$ colors (hence in the predicted
$(B-V)_{0,2M}$ colors), as well as in the reddening estimates, there will be stars with
incorrect assignments of BHB likelihood, but we believe that the exercise is
still of value.  

If a star has no measured $B-V$ color, and no 2MASS information is available,
which can occur for a variety of reasons, including the possibility that the
star is an RR Lyrae variable caught by 2MASS at the ``wrong'' (i.e.,
faint) phase, or has poor coordinates from the HK-survey prism plates,
the above assignments of BHB likelihood cannot be made. As discussed in \S
\ref{match2mass}, some candidates have multiple possible 2MASS
identifications that could not be resolved with confidence, or had flags raised
in the 2MASS catalog that indicated possible contamination from nearby stars or
other problems.  These are identified in the catalog listing discussed below.

\section {The Catalog}

Table 3 presents our final catalog of FHB/A candidates assembled from the
HK survey. We have grouped the stars by the plate on which they were originally
discovered. 

For each FHB/A candidate, Table 3 lists: The star name (Column 1), its
brightness classification from visual inspection of the HK survey plates (Column
2), its coordinates from the HK-survey measurements (Columns 3 and 4), its
Galactic coordinates (Columns 5 and 6), its measured $V$ magnitude (Column 7)
and $B-V$ color (Column 8), where available, the adopted reddening estimate
(Column 9), its 2MASS $J$ magnitude (Column 10) and 2MASS $J-H$ color (Column
11), where available, its extinction-corrected $V_0$ magnitude (Column 12), its
predicted $V_{0,2M}$ magnitude (Column 13), based on the correlations discussed
above, its reddening-corrected $(B-V)_0$ color (Column 14), and its predicted
$(B-V) _{0,2M}$ color (Column 15). The BHB probability classifications discussed
above are listed in Column 16. Stars where there remained some difficulty in
either assigning a confident 2MASS match, or where the 2MASS flags indicated
possible problems with the photometry, are indicated by the ``:'' attached to
the probability classification. Stars for which no suitable match in 2MASS could
be found are assigned an ``X'' for their probability classifications. In Column
17 we list, where available, the classification of the star assigned by Wilhelm
et al. (1999b), which requires that the star have both $UBV$ photometry and
medium-resolution spectroscopy available. These classifications are: (1) FHB --
indicating that the star has a derived temperature and gravity consistent with a
horizontal-branch identification, (2) FHB/A -- indicating that it was not
possible, based on the data in hand, to differentiate between a
horizontal-branch and a high-gravity A-type star identification, (3) A-V --
indicating that the star is a likely high-gravity object (note that this
includes some stars with temperatures cooler than would usually be considered
for assignment of a type A designation), and (4) Am -- indicating that the star
is a likely metallic-line A-type star. The interested reader should consult
Wilhelm et al. (1999a) for a more detailed discussion of these classes.
  
Table 4 lists the adopted 2MASS identifications for the entire set of stars
listed in Table 3, as well as their 2MASS
coordinates (which are in most cases more accurate than the HK-survey
coordinates listed in Table 3).  In the final column the stars with multiple
HK-survey identifications are provided. Note that some of these
multiply-identified stars are not present in the FHB/A candidate catalog, since
they were not classified as such in the HK survey.

\section{Discussion}

From the sample containing 11453 (unique) FHB/A candidates with available
measurements or predictions of (B-V)$_0$, 7250 (63\%) are assigned to the
high-likelihood BHB category, 2964 (26\%) to the medium-likelihood BHB category,
and 1239 (11\%) to the low-likelihood BHB category.  

The combination of a low-resolution spectroscopic assignment based on the
presence of a broad H-$\epsilon$ feature in the HK-survey prism spectrum, and
colors (either measured or estimated from 2MASS) should result in a relatively
clean sample of BHB candidates. As mentioned above, this still does not remove
the difficulty that is expected to arise from the presence of blue stragglers in
the same color range as BHB candidates. However, at least it should practically
eliminate potential contamination from stars that may be too hot or too cool to
be likely BHB stars, and thus should enable far more efficient spectroscopic
follow-up of this sample for a variety of purposes.

We can obtain an idea of the success of the likelihood assignments from
consideration of the classifications given for the subset of 661 unique stars
included in our present catalog that have been studied by Wilhelm et al. (1999b).
Table 5 summarizes this comparison. As can be seen from the Table, 78\% of the
unique candidates in common with Wilhelm et al. for which we assign likelihood
class ``H'' were classified by Wilhelm et al. as FHB or FHB/A (their
indeterminant gravity class). For the likelihood classes ``M'' and ``L'', these
fractions drop to 37\% and 30\%, respectively. We conclude that future
observations of the likelihood class ``H'' stars will contain a large fraction
of bona-fide BHB stars. Many of the remaining stars in this class will be likely
blue stragglers.

The likely BHB stars in our catalog form the basis for extensive studies of the
kinematics and structure of the thick disk and inner halo of the Galaxy. In
fact, this has already begun. Over the past few years, we have been gathering
medium-resolution spectroscopy for many hundreds of the BHB targets listed in
Table 3, in particular in the fields of greatest interest for assessment of the
dynamical model described in Sommer-Larsen et al. (1997). These data will be
combined with additional observations of HES FHB candidates, which reach to
greater distances than the HK survey candidates, in order to better constrain
the model. Furthermore, the high-likelihood BHB candidates that are located in
the directions of the high velocity clouds of H~I (Wakker \& van Woerden 1997)
are being studied at high spectral resolution in order to bracket the distances
to such clouds (see, e.g., Thom et al. 2006). Since many of the BHB candidates
in our catalog are expected to be located within 10 kpc of the Sun, it is
possible to make use of detectable proper motions (either alone, or in
combination with available radial velocities in order to obtain full space
motions) that are either already available from a number of sources (e.g.,
UCAC2; Zacharias et al. 2004; USNOB+SDSS; Munn et al. 2004) or that will become
available in the near future, e.g., from ongoing surveys such as the Southern
Proper Motion program of van Altena and colleagues (Girard et al. 2004 ). Such
information will provide valuable constraints on the change in the kinematics of
the Galaxy in the poorly understood thick-disk/halo transition, and can be
combined with other much more distant samples of BHB stars (e.g., from SDSS, see
Sakamoto et al. 2003, Sirko et al. 2004b) in order to study the distribution of
mass from the inner to the outer halo of the Galaxy. 

\acknowledgements

T.C.B. and B.M. acknowledge partial support from a series of grants awarded by the US
National Science Foundation which made the HK survey possible, as well as from grant
PHY 02-16783; Physics Frontier Center/Joint Institute for Nuclear Astrophysics
(JINA). S.R. and T.A. acknowledge partial support from CNPq, FAPESP, and Capes.

\begin{deluxetable}{lccrrcc}
\tablecolumns{7}
\tablewidth{0pt}
\tabletypesize{\scriptsize}
\tablenum{1}
\tablecaption{HK-Survey Objective-Prism Survey Plate Log}
\tablehead
{
\colhead{Plate}  & \multicolumn{2}{c}{RA (2000) DEC} &  \colhead{l}    &
\colhead{b}      & \colhead{N (FHB/A)}               & \colhead{Repeats}
}
\startdata
BS~15620 & 14:59.5 &$+$44:04 &  74.2 & $+$59.0 &  26&\\
BS~15621 & 10:21.7 &$+$25:24 & 207.9 & $+$57.1 &  24&\\
BS~15622 & 12:55.0 &$+$25:52 & 344.7 & $+$88.0 &  22&\\
BS~15623 & 14:01.1 &$+$25:47 &  31.4 & $+$73.8 &  17&       BS~16023\\
BS~15624 & 16:17.9 &$+$45:36 &  71.3 & $+$45.2 &  34&\\
BS~15625 & 11:53.4 &$+$25:41 & 217.6 & $+$77.4 &   8&\\
BS~15626 & 13:37.0 &$+$24:29 &  20.6 & $+$78.8 &  20&\\ 
BS~15627 & 15:34.7 &$+$49:07 &  78.8 & $+$51.6 &  25&\\
BS~16022 & 12:10.8 &$+$25:33 & 223.0 & $+$81.3 &  21&\\
BS~16023 & 14:01.9 &$+$24:31 &  27.1 & $+$73.4 &  26&       BS~15623\\
BS~16026 & 12:25.7 &$+$29:16 & 191.5 & $+$84.5 &  16&\\
BS~16027 & 13:12.2 &$+$30:37 &  66.9 & $+$84.0 &  29&\\
BS~16029 & 16:06.7 &$+$50:00 &  78.0 & $+$46.4 &  51&\\
BS~16031 & 12:33.9 &$+$25:15 & 244.4 & $+$86.0 &  29&\\
BS~16032 & 12:50.6 &$+$30:10 & 116.0 & $+$87.2 &  43&\\
BS~16033 & 13:16.4 &$+$24:05 &   6.4 & $+$83.0 &  40&\\
BS~16034 & 15:46.8 &$+$55:19 &  86.8 & $+$47.6 &  49&\\
BS~16076 & 13:02.3 &$+$19:33 & 325.0 & $+$81.6 &  16&\\
BS~16077 & 11:46.7 &$+$30:08 & 198.0 & $+$76.1 &  22&\\
BS~16078 & 13:40.4 &$+$30:04 &  49.4 & $+$78.4 &  32&\\
BS~16079 & 16:00.8 &$+$59:29 &  91.2 & $+$44.3 &  46&\\
BS~16080 & 16:49.5 &$+$59:38 &  89.0 & $+$38.4 &  43&\\
BS~16081 & 12:03.3 &$+$30:21 & 193.9 & $+$79.6 &  32&\\
BS~16082 & 13:56.4 &$+$29:47 &  46.2 & $+$75.0 &  64&\\
BS~16083 & 15:00.0 &$+$49:23 &  83.2 & $+$56.6 &  39&\\
BS~16084 & 16:16.5 &$+$55:13 &  84.6 & $+$43.7 &  44&\\
BS~16085 & 12:35.9 &$+$19:58 & 280.3 & $+$82.0 &  28&\\
BS~16086 & 15:22.7 &$+$59:40 &  94.8 & $+$48.5 &  29&\\
BS~16087 & 16:20.9 &$+$64:44 &  96.6 & $+$40.1 &  34&\\
BS~16088 & 16:46.2 &$+$45:04 &  70.3 & $+$40.2 &  56&\\
BS~16089 & 13:57.7 &$+$35:22 &  65.9 & $+$73.3 &  29&\\
BS~16090 & 14:27.9 &$+$51:07 &  91.5 & $+$59.8 &  26&\\
BS~16466 & 12:48.4 &$+$33:30 & 124.2 & $+$83.9 &  27&\\
BS~16467 & 13:44.1 &$+$20:31 &   7.3 & $+$75.7 &  45&\\
BS~16468 & 09:15.0 &$+$39:41 & 183.1 & $+$44.4 &  39&\\
BS~16469 & 10:05.1 &$+$40:53 & 180.3 & $+$53.9 &  23&\\
BS~16470 & 12:17.3 &$+$15:20 & 270.0 & $+$75.9 &  39&\\
BS~16471 & 13:58.9 &$+$40:42 &  80.2 & $+$70.4 &  39&\\
BS~16472 & 15:03.6 &$-$00:16 & 358.0 & $+$47.5 &  45&       CS~30301\\
BS~16473 & 08:52.7 &$+$44:44 & 176.1 & $+$40.4 &  40&\\
BS~16474 & 10:35.3 &$+$39:24 & 180.9 & $+$59.8 &  27&\\
BS~16477 & 14:44.2 &$+$05:03 & 359.0 & $+$54.6 &  45&       CS~30317\\
BS~16478 & 11:05.8 &$+$40:40 & 174.3 & $+$65.0 &  26&       BS~16553\\
BS~16479 & 13:21.5 &$+$20:13 & 350.8 & $+$79.7 &  39&       BS~16543\\
BS~16541 & 15:23.4 &$+$08:43 &  13.3 & $+$49.1 &  54&\\
BS~16542 & 16:32.4 &$+$21:09 &  38.9 & $+$39.1 &  43&\\
BS~16543 & 13:22.5 &$+$20:33 & 353.0 & $+$79.7 &  34&       BS~16479\\
BS~16545 & 11:14.5 &$+$36:25 & 182.8 & $+$68.1 &  35&\\
BS~16546 & 14:21.2 &$+$09:57 & 358.6 & $+$62.2 &  42&       CS~22883\\
BS~16547 & 15:23.4 &$-$04:17 & 358.7 & $+$41.1 &  46&\\
BS~16548 & 16:38.6 &$+$50:27 &  77.5 & $+$41.3 &  64&\\
BS~16549 & 12:00.4 &$+$35:54 & 170.5 & $+$76.6 &  24&\\
BS~16550 & 14:02.6 &$+$15:48 &   2.6 & $+$69.4 &  37&\\
BS~16551 & 15:23.4 &$-$08:44 & 354.5 & $+$38.0 &  74&\\
BS~16552 & 16:58.7 &$+$34:53 &  57.6 & $+$36.9 &  74&\\
BS~16553 & 10:58.3 &$+$39:44 & 177.6 & $+$64.0 &  18&       BS~16478\\
BS~16554 & 14:02.3 &$+$20:18 &  14.0 & $+$71.8 &  34&\\
BS~16555 & 15:01.9 &$-$04:55 & 352.9 & $+$44.5 &  36&\\
BS~16556 & 16:44.8 &$+$29:38 &  50.3 & $+$38.7 &  45&\\
BS~16557 & 11:19.4 &$-$08:32 & 268.7 & $+$48.0 &  29&\\
BS~16558 & 12:21.9 &$-$10:04 & 292.1 & $+$52.0 &  16&\\
BS~16559 & 15:20.2 &$+$00:37 &   3.0 & $+$44.9 &  51&       CS~22890\\
BS~16920 & 12:18.0 &$+$40:34 & 146.9 & $+$75.4 &  22&\\
BS~16921 & 14:31.2 &$+$45:41 &  81.9 & $+$62.7 &  30&\\
BS~16922 & 08:50.6 &$+$50:10 & 169.0 & $+$39.9 &  41&\\
BS~16923 & 10:16.8 &$+$45:44 & 171.3 & $+$54.8 &  56&\\
BS~16924 & 13:39.1 &$+$35:52 &  74.7 & $+$76.4 &  22&\\
BS~16926 & 09:19.2 &$+$45:33 & 174.8 & $+$45.0 &  49&\\
BS~16927 & 09:41.3 &$+$40:03 & 182.4 & $+$49.5 &  34&\\
BS~16928 & 12:22.2 &$+$34:59 & 158.4 & $+$80.5 &  17&\\
BS~16929 & 13:13.6 &$+$35:40 &  92.2 & $+$80.2 &  23&\\
BS~16930 & 14:35.7 &$+$55:27 &  95.9 & $+$56.0 &  24&\\
BS~16933 & 12:13.9 &$+$20:07 & 253.3 & $+$79.2 &  18&\\
BS~16934 & 13:40.0 &$+$15:53 & 351.4 & $+$73.5 &  31&\\
BS~16935 & 09:53.0 &$+$45:25 & 173.6 & $+$50.9 &  36&\\
BS~16936 & 11:54.1 &$+$19:59 & 240.6 & $+$75.4 &  21&\\
BS~16938 & 13:07.3 &$+$39:46 & 107.1 & $+$77.1 &  18&\\
BS~16939 & 15:43.3 &$+$09:52 &  18.5 & $+$45.4 &  55&\\
BS~16940 & 09:40.0 &$+$34:59 & 190.2 & $+$49.3 &  35&\\
BS~16941 & 11:53.1 &$+$40:35 & 160.6 & $+$72.5 &  22&\\
BS~16942 & 12:54.9 &$+$15:43 & 310.0 & $+$78.3 &  23&\\
BS~16945 & 10:47.7 &$+$35:51 & 187.2 & $+$63.0 &  26&       BS~17137\\
BS~16968 & 15:00.7 &$+$05:56 &   4.6 & $+$52.1 &  35&       CS~30325\\
BS~16972 & 13:19.8 &$+$16:02 & 336.7 & $+$76.6 &  13&\\
BS~16981 & 14:41.4 &$+$00:50 & 353.2 & $+$52.2 &  45&\\
BS~16984 & 12:38.7 &$+$15:12 & 291.4 & $+$77.6 &  26&\\
BS~16986 & 12:01.4 &$+$09:34 & 268.5 & $+$68.9 &  28&\\
BS~16990 & 12:22.8 &$+$10:17 & 281.9 & $+$71.8 &  35&\\
BS~17135 & 09:21.3 &$+$54:42 & 162.1 & $+$43.8 &  28&\\
BS~17136 & 08:50.6 &$+$36:08 & 187.3 & $+$39.4 &  29&\\
BS~17137 & 10:49.4 &$+$35:11 & 188.5 & $+$63.4 &  12&       BS~16945\\
BS~17138 & 14:29.2 &$+$35:08 &  59.2 & $+$67.3 &  18&\\
BS~17139 & 09:02.7 &$+$29:22 & 196.6 & $+$40.7 &  35&\\
BS~17140 & 11:24.3 &$+$39:18 & 173.5 & $+$68.8 &  23&\\
BS~17141 & 12:42.2 &$+$39:45 & 129.1 & $+$77.6 &  18&\\
BS~17142 & 10:23.5 &$+$35:02 & 190.0 & $+$58.2 &  18&\\
BS~17435 & 11:55.3 &$+$15:11 & 254.7 & $+$72.5 &  10&\\
BS~17436 & 13:33.7 &$+$44:59 &  99.1 & $+$70.3 &  22&\\
BS~17438 & 07:55.6 &$+$40:24 & 179.9 & $+$29.5 &  48&\\
BS~17439 & 10:54.2 &$-$14:26 & 265.8 & $+$39.7 &  22&\\
BS~17440 & 14:00.8 &$+$49:38 &  96.1 & $+$63.9 &  23&\\
BS~17444 & 08:23.4 &$+$40:05 & 181.4 & $+$34.6 &  26&\\
BS~17446 & 15:28.5 &$+$45:41 &  74.0 & $+$53.6 &  29&\\
BS~17447 & 18:05.0 &$+$59:56 &  88.8 & $+$28.9 &  48&\\
BS~17448 & 08:49.2 &$+$40:01 & 182.2 & $+$39.5 &  16&\\
BS~17449 & 12:05.0 &$-$09:41 & 285.5 & $+$51.4 &  10&\\
BS~17450 & 13:34.9 &$+$39:49 &  88.5 & $+$74.4 &  13&\\
BS~17451 & 15:41.7 &$+$64:03 &  98.3 & $+$44.2 &  21&\\
BS~17569 & 22:03.9 &$+$05:48 &  66.5 & $-$38.1 &  26&\\
BS~17570 & 00:18.0 &$+$25:16 & 114.2 & $-$36.8 &  31&\\
BS~17571 & 04:21.6 &$+$10:26 & 184.2 & $-$26.1 &  60&\\
BS~17572 & 09:22.4 &$-$04:33 & 237.4 & $+$30.8 &  36&\\
BS~17574 & 04:41.9 &$+$10:36 & 187.3 & $-$22.0 &  73&\\
BS~17575 & 05:01.6 &$+$10:42 & 190.2 & $-$18.0 &  93&\\
BS~17576 & 09:40.5 &$-$04:53 & 241.0 & $+$34.2 &  31&\\
BS~17577 & 09:41.1 &$-$09:21 & 245.2 & $+$31.5 &  32&\\
BS~17578 & 21:42.8 &$+$15:14 &  70.6 & $-$27.9 &  39&\\
BS~17579 & 00:53.6 &$+$01:29 & 125.4 & $-$61.1 &  13&\\
BS~17580 & 04:00.1 &$+$15:31 & 176.1 & $-$26.8 &  43&\\
BS~17581 & 09:18.7 &$-$10:15 & 242.0 & $+$26.7 &  56&\\
BS~17582 & 10:29.6 &$-$15:00 & 260.1 & $+$35.8 &  39&\\
BS~17583 & 21:42.4 &$+$25:28 &  78.5 & $-$20.6 &  73&\\
BS~17584 & 23:32.2 &$+$25:15 & 101.9 & $-$34.2 &  32&\\
BS~17585 & 04:22.4 &$+$15:14 & 180.2 & $-$23.0 &  68&\\
BS~17586 & 09:01.0 &$-$09:54 & 238.8 & $+$23.4 &  59&\\
BS~17587 & 09:51.0 &$-$15:03 & 252.0 & $+$29.5 &  42&\\
CS~22166 & 01:04.4 &$-$13:04 & 137.8 & $-$75.2 &  14&\\
CS~22167 & 03:17.1 &$-$05:11 & 187.5 & $-$48.4 &  13&\\
CS~22169 & 04:08.6 &$-$14:43 & 208.3 & $-$42.1 &  10&\\
CS~22170 & 00:44.9 &$-$09:57 & 119.6 & $-$72.5 &  16&\\
CS~22171 & 02:01.5 &$-$09:44 & 170.5 & $-$65.4 &  14&\\
CS~22172 & 03:24.8 &$-$09:49 & 195.0 & $-$49.4 &  16&\\
CS~22173 & 04:07.0 &$-$18:46 & 213.2 & $-$44.0 &  18&       CS~30494\\
CS~22174 & 01:22.0 &$-$10:27 & 149.2 & $-$71.4 &  18&\\
CS~22175 & 02:21.6 &$-$09:31 & 178.1 & $-$61.8 &  13&\\
CS~22176 & 03:43.7 &$-$10:07 & 198.8 & $-$45.5 &  12&\\
CS~22177 & 04:14.6 &$-$24:59 & 222.2 & $-$44.1 &  18&\\
CS~22179 & 00:42.3 &$-$04:37 & 118.7 & $-$67.2 &  13&\\
CS~22180 & 01:37.4 &$-$10:14 & 158.9 & $-$69.4 &  14&\\
CS~22181 & 03:02.0 &$-$11:04 & 192.0 & $-$54.8 &  14&\\
CS~22182 & 04:18.5 &$-$29:44 & 228.9 & $-$44.3 &  26&\\
CS~22183 & 00:58.6 &$-$04:23 & 129.1 & $-$66.9 &  13&\\
CS~22184 & 02:40.8 &$-$09:38 & 184.4 & $-$58.2 &  17&\\
CS~22185 & 03:23.7 &$-$14:19 & 201.0 & $-$51.8 &   9&\\
CS~22186 & 04:22.7 &$-$34:43 & 236.0 & $-$44.2 &  21&\\
CS~22188 & 00:49.6 &$-$38:13 & 302.4 & $-$79.2 &  21&\\
CS~22189 & 02:45.0 &$-$15:02 & 194.3 & $-$60.4 &  16&\\
CS~22190 & 03:48.1 &$-$14:24 & 205.0 & $-$46.5 &  21&\\
CS~22191 & 04:35.3 &$-$39:36 & 243.0 & $-$42.1 &  11&\\
CS~22871 & 14:40.8 &$-$19:55 & 335.8 & $+$35.5 &  74&\\
CS~22872 & 16:25.8 &$-$03:20 &  11.4 & $+$29.5 &  98&\\
CS~22873 & 20:01.8 &$-$59:39 & 337.7 & $-$32.5 & 123&\\
CS~22874 & 14:40.4 &$-$25:05 & 332.6 & $+$31.0 &  97&\\
CS~22875 & 22:31.0 &$-$39:52 &   0.4 & $-$58.9 &  30&\\
CS~22876 & 15:11.0 &$-$34:40 & 333.4 & $+$19.4 &  30&\\
CS~22877 & 13:20.4 &$-$10:24 & 315.4 & $+$51.4 &  43&\\
CS~22878 & 16:41.5 &$+$09:28 &  26.6 & $+$32.6 &  90&\\
CS~22879 & 20:47.8 &$-$40:03 &   2.2 & $-$39.3 &  92&\\
CS~22880 & 20:45.6 &$-$19:37 &  27.0 & $-$34.2 &  91&\\
CS~22881 & 22:06.6 &$-$40:03 &   1.6 & $-$54.3 &  39&\\
CS~22882 & 00:29.2 &$-$29:46 &   3.2 & $-$85.0 &  17&\\
CS~22883 & 14:20.0 &$+$10:13 & 358.6 & $+$62.5 &  27&       BS~16546\\
CS~22884 & 15:41.4 &$-$10:04 & 357.2 & $+$33.9 &  95&\\
CS~22885 & 20:22.6 &$-$39:40 &   1.9 & $-$34.4 & 136&\\
CS~22886 & 22:19.8 &$-$09:24 &  52.9 & $-$50.5 &  41&       CS~29512\\
CS~22887 & 22:41.7 &$-$09:10 &  58.1 & $-$54.9 &  27&\\
CS~22888 & 23:11.4 &$-$35:00 &   6.8 & $-$67.9 &  33&       CS~30493\\
CS~22889 & 13:42.7 &$-$10:13 & 323.9 & $+$50.2 &  44&\\
CS~22890 & 15:19.6 &$+$01:19 &   3.6 & $+$45.5 &  69&       BS~16559\\
CS~22891 & 19:20.1 &$-$59:43 & 337.2 & $-$27.3 & 157&\\
CS~22892 & 22:08.5 &$-$15:10 &  42.8 & $-$50.9 &  29&\\
CS~22893 & 23:02.1 &$-$09:27 &  63.3 & $-$59.1 &  22&\\
CS~22894 & 23:39.2 &$-$00:21 &  88.3 & $-$58.2 &  21&\\
CS~22896 & 19:34.9 &$-$54:50 & 343.0 & $-$28.5 & 143&\\
CS~22897 & 21:21.1 &$-$64:23 & 329.7 & $-$40.6 &  72&\\
CS~22898 & 21:07.4 &$-$19:58 &  28.7 & $-$39.1 &  54&\\
CS~22936 & 18:54.0 &$-$34:32 &   2.0 & $-$16.1 & 239&\\
CS~22937 & 21:14.2 &$-$39:47 &   2.9 & $-$44.3 &  60&       CS~30492\\
CS~22938 & 22:57.0 &$-$64:57 & 320.5 & $-$48.5 &  38&\\
CS~22939 & 19:29.6 &$-$29:36 &   9.7 & $-$21.3 & 183&\\
CS~22940 & 20:40.7 &$-$59:58 & 336.8 & $-$37.3 &  74&\\
CS~22941 & 23:35.8 &$-$35:00 &   2.5 & $-$72.6 &  25&\\
CS~22942 & 00:54.6 &$-$23:55 & 143.3 & $-$86.3 &   8&\\
CS~22943 & 20:22.7 &$-$45:01 & 355.4 & $-$35.1 & 126&\\
CS~22944 & 21:46.1 &$-$14:48 &  39.9 & $-$45.8 &  41&\\
CS~22945 & 21:42.4 &$-$65:07 & 327.5 & $-$42.4 &  48&\\
CS~22946 & 01:20.4 &$-$18:59 & 165.3 & $-$78.9 &   7&\\
CS~22947 & 19:18.3 &$-$49:46 & 348.0 & $-$25.1 & 193&\\
CS~22948 & 21:42.3 &$-$39:37 &   3.0 & $-$49.7 &  63&\\
CS~22949 & 23:21.5 &$-$04:29 &  76.8 & $-$59.0 &  23&\\
CS~22950 & 20:26.7 &$-$14:44 &  30.3 & $-$28.2 & 121&\\
CS~22951 & 21:52.5 &$-$44:41 & 354.8 & $-$51.0 &  95&\\
CS~22952 & 23:39.2 &$-$03:48 &  84.7 & $-$61.1 &  17&\\
CS~22953 & 01:20.4 &$-$60:11 & 295.9 & $-$56.8 &  32&\\
CS~22954 & 02:41.5 &$-$05:03 & 178.2 & $-$55.1 &  10&\\
CS~22955 & 20:30.9 &$-$25:13 &  19.3 & $-$32.8 & 106&\\
CS~22956 & 22:06.2 &$-$64:34 & 326.2 & $-$44.8 &  61&\\
CS~22957 & 15:14.9 &$-$04:24 & 356.6 & $+$42.6 &  11&\\
CS~22958 & 01:58.1 &$-$54:53 & 283.2 & $-$59.8 &  19&\\
CS~22959 & 18:58.7 &$-$65:15 & 330.6 & $-$25.7 & 163&\\
CS~22961 & 08:05.4 &$-$30:07 & 248.2 & $+$ 1.2 &  21&\\
CS~22962 & 01:41.5 &$-$05:00 & 154.5 & $-$64.4 &   8&\\
CS~22963 & 03:03.5 &$-$04:47 & 183.9 & $-$50.8 &  20&\\
CS~22964 & 19:56.3 &$-$39:38 &   0.8 & $-$29.4 & 158&\\
CS~22965 & 22:02.5 &$-$05:04 &  54.8 & $-$44.6 &  42&\\
CS~22966 & 23:44.7 &$-$29:57 &  18.9 & $-$75.7 &  22&       CS~29496\\
CS~22967 & 01:20.4 &$-$04:42 & 142.7 & $-$66.1 &  16&\\
CS~22968 & 03:13.8 &$-$54:25 & 268.8 & $-$52.3 &  26&\\
CS~29491 & 22:36.4 &$-$30:18 &  19.3 & $-$60.9 &  35&\\
CS~29493 & 21:49.0 &$-$30:17 &  17.7 & $-$50.7 &  46&\\
CS~29494 & 23:21.1 &$-$29:49 &  20.6 & $-$70.6 &  26&\\
CS~29495 & 21:42.4 &$-$25:36 &  24.4 & $-$48.5 &  39&\\
CS~29496 & 23:43.3 &$-$30:07 &  18.4 & $-$75.4 &  22&       CS~22966\\
CS~29497 & 00:36.1 &$-$24:40 &  75.5 & $-$86.0 &  12&\\
CS~29498 & 21:00.1 &$-$29:32 &  16.2 & $-$40.2 &  62&\\
CS~29499 & 23:45.3 &$-$24:52 &  39.4 & $-$75.5 &  31&\\
CS~29500 & 02:01.5 &$-$30:34 & 228.7 & $-$73.9 &  15&\\
CS~29501 & 21:15.4 &$-$34:59 &   9.6 & $-$44.2 &  51&\\
CS~29502 & 22:20.5 &$+$00:22 &  64.7 & $-$44.8 &  24&\\
CS~29503 & 00:09.2 &$-$24:47 &  47.0 & $-$80.7 &  11&\\
CS~29504 & 01:36.9 &$-$34:00 & 248.7 & $-$77.8 &  14&\\
CS~29505 & 03:26.4 &$-$45:24 & 253.7 & $-$54.1 &  29&\\
CS~29506 & 21:27.0 &$-$19:51 &  30.9 & $-$43.4 &  47&\\
CS~29509 & 00:53.2 &$-$30:18 & 285.6 & $-$87.0 &  14&\\
CS~29510 & 02:20.1 &$-$24:52 & 211.1 & $-$69.4 &  24&\\
CS~29512 & 22:17.9 &$-$10:16 &  51.3 & $-$50.6 &  30&       CS~22886\\
CS~29513 & 23:26.0 &$-$40:10 & 351.0 & $-$68.6 &  16&\\
CS~29514 & 01:13.4 &$-$24:28 & 186.3 & $-$83.8 &  20&\\
CS~29515 & 02:43.0 &$-$30:38 & 227.3 & $-$64.9 &  32&\\
CS~29516 & 22:20.6 &$+$04:25 &  68.8 & $-$42.1 &  42&\\
CS~29517 & 17:52.8 &$-$15:18 &  12.9 & $+$ 4.9 &  12&\\
CS~29518 & 01:16.3 &$-$29:59 & 238.0 & $-$83.5 &  16&\\
CS~29519 & 02:32.9 &$-$50:24 & 269.4 & $-$59.7 &  28&\\
CS~29520 & 04:36.4 &$-$60:15 & 270.3 & $-$39.8 &  27&\\
CS~29521 & 22:59.5 &$+$09:41 &  83.6 & $-$44.4 &  24&\\
CS~29522 & 23:43.1 &$+$10:04 &  97.7 & $-$49.3 &  17&       CS~30333\\
CS~29526 & 03:31.0 &$-$25:33 & 219.5 & $-$53.9 &  21&\\
CS~29527 & 00:36.4 &$-$20:16 & 100.4 & $-$82.3 &  21&\\
CS~29528 & 02:23.8 &$-$19:41 & 198.5 & $-$67.0 &  14&\\
CS~29529 & 03:58.2 &$-$60:22 & 272.5 & $-$44.2 &  26&\\
CS~30301 & 15:01.0 &$+$00:21 & 358.1 & $+$48.4 &  58&       BS~16472\\
CS~30302 & 19:48.1 &$-$50:09 & 348.6 & $-$29.9 & 118&\\
CS~30303 & 21:24.5 &$-$30:22 &  16.4 & $-$45.5 &  82&\\
CS~30306 & 15:24.3 &$+$05:26 &   9.4 & $+$47.1 &  67&\\
CS~30308 & 20:52.6 &$-$45:05 & 355.7 & $-$40.4 &  66&\\
CS~30310 & 01:09.4 &$-$39:22 & 284.8 & $-$77.3 &  29&\\
CS~30311 & 13:18.4 &$+$04:30 & 321.3 & $+$66.1 &  32&\\
CS~30312 & 15:41.1 &$-$00:04 &   6.7 & $+$40.4 &  94&\\
CS~30314 & 20:47.9 &$-$34:50 &   8.8 & $-$38.6 &  73&\\
CS~30315 & 23:27.4 &$-$25:06 &  35.5 & $-$71.6 &  34&\\
CS~30317 & 14:38.9 &$+$04:32 & 356.8 & $+$55.2 &  42&       BS~16477\\
CS~30319 & 21:11.4 &$-$25:09 &  22.6 & $-$41.6 &  38&       CS~30331\\
CS~30320 & 14:00.2 &$+$10:12 & 350.7 & $+$66.0 &  51&\\
CS~30322 & 21:19.5 &$-$45:07 & 355.4 & $-$45.1 &  88&\\
CS~30323 & 23:05.5 &$-$25:09 &  32.6 & $-$66.7 &  34&\\
CS~30324 & 00:16.3 &$-$19:30 &  77.7 & $-$79.1 &  28&\\
CS~30325 & 14:59.0 &$+$05:20 &   3.4 & $+$52.0 &  31&       BS~16968\\
CS~30327 & 21:59.0 &$-$34:59 &  10.3 & $-$53.1 &  41&\\
CS~30329 & 15:39.1 &$-$05:11 &   1.1 & $+$37.6 &  59&\\
CS~30331 & 21:13.9 &$-$24:52 &  23.2 & $-$42.0 &  71&       CS~30319\\
CS~30332 & 22:41.9 &$+$10:26 &  79.5 & $-$41.2 &  41&\\
CS~30333 & 23:41.7 &$+$09:55 &  97.1 & $-$49.3 &  30&       CS~29522\\
CS~30336 & 20:39.7 &$-$30:13 &  14.1 & $-$36.0 & 100&\\
CS~30337 & 22:09.0 &$-$30:03 &  18.9 & $-$55.0 &  48&\\
CS~30338 & 23:22.8 &$+$09:24 &  90.4 & $-$47.7 &  36&\\
CS~30343 & 21:40.9 &$-$35:01 &  10.1 & $-$49.4 &  69&\\
CS~30344 & 22:51.1 &$-$35:24 &   7.9 & $-$63.7 &  42&\\
CS~30492 & 21:12.8 &$-$39:48 &   2.9 & $-$44.0 &  73&       CS~22937\\
CS~30493 & 23:11.7 &$-$35:04 &   6.6 & $-$67.9 &  33&       CS~22888\\
CS~30494 & 04:05.7 &$-$19:11 & 213.6 & $-$44.4 &  19&       CS~22173\\
CS~31060 & 00:11.2 &$-$15:09 &  85.4 & $-$74.9 &  27&\\
CS~31061 & 02:34.6 &$+$04:35 & 165.6 & $-$49.1 &  28&       CS~31063\\
CS~31062 & 00:38.0 &$-$14:15 & 111.4 & $-$76.6 &  17&\\
CS~31063 & 02:37.5 &$+$04:35 & 166.5 & $-$48.7 &  30&       CS~31061\\
CS~31064 & 02:48.4 &$-$65:34 & 285.3 & $-$47.5 &  52&\\
CS~31065 & 00:55.1 &$-$14:23 & 129.6 & $-$76.9 &  12&\\
CS~31066 & 01:56.2 &$-$20:03 & 191.1 & $-$73.0 &  19&\\
CS~31067 & 03:17.5 &$+$04:55 & 176.7 & $-$41.7 &  25&\\
CS~31068 & 05:52.8 &$-$59:56 & 268.7 & $-$30.4 &  53&\\
CS~31069 & 00:17.5 &$+$05:36 & 108.8 & $-$56.1 &  27&       CS~31070\\
CS~31070 & 00:18.2 &$+$04:49 & 108.9 & $-$56.9 &  31&       CS~31079\\
CS~31071 & 03:17.1 &$+$15:12 & 167.7 & $-$34.3 &  53&\\
CS~31072 & 05:24.4 &$-$59:02 & 267.6 & $-$34.0 &  63&       CS~31076\\
CS~31074 & 02:45.0 &$+$09:58 & 163.9 & $-$43.2 &  37&\\
CS~31075 & 03:31.2 &$-$65:24 & 280.8 & $-$44.4 &  40&\\
CS~31076 & 05:24.1 &$-$59:03 & 267.7 & $-$34.1 &  40&       CS~31072\\
CS~31077 & 00:12.9 &$-$40:56 & 329.4 & $-$74.6 &  30&\\
CS~31078 & 02:57.3 &$+$06:04 & 170.6 & $-$44.4 &  34&       CS~31079\\
CS~31079 & 02:57.5 &$+$05:53 & 170.9 & $-$44.5 &  32&       CS~31078\\
CS~31080 & 04:49.2 &$-$45:21 & 250.7 & $-$39.7 &  55&\\
CS~31081 & 01:22.0 &$-$13:52 & 154.3 & $-$74.4 &  25&       CS~31082\\
CS~31082 & 01:21.8 &$-$14:14 & 154.8 & $-$74.8 &  34&       CS~31081\\
CS~31083 & 03:24.9 &$+$10:02 & 173.7 & $-$36.8 &  36&       CS~31084\\
CS~31084 & 03:24.8 &$+$09:41 & 174.0 & $-$37.1 &  46&       CS~31083\\
CS~31085 & 07:12.3 &$+$09:45 & 206.9 & $+$ 9.6 &  37&\\
CS~31086 & 01:39.4 &$-$20:03 & 183.2 & $-$76.3 &  16&\\
CS~31087 & 03:39.7 &$+$04:59 & 181.5 & $-$37.6 &  45&\\
CS~31088 & 23:42.9 &$+$04:24 &  93.8 & $-$54.4 &  37&\\
CS~31089 & 01:35.0 &$-$15:51 & 168.1 & $-$74.3 &  28&       CS~31090\\
CS~31090 & 01:34.4 &$-$15:58 & 168.0 & $-$74.5 &  34&       CS~31089\\
\enddata
\end{deluxetable}

\clearpage

\begin{deluxetable}{lccccccc}
\tablecolumns{8}
\tablenum{2}
\tablewidth{0pt}
\tablecaption{2MASS $J$ magnitudes and Robust One-Sigma Scatter
in this Magnitude Associated with Each Brightness Class}
\tablehead
{
\colhead{Estimator} &\colhead{BC 1} & \colhead{BC 2} & \colhead{BC 3} &
\colhead{BC 4} & \colhead{BC 5} & \colhead{BC 6} & \colhead{BC 7}
}
\startdata
Central Location & 11.567 & 12.013 & 12.244 & 13.142 & 13.546 & 14.138 & 14.617 \\
Scale            & 1.018  & 1.194  & 1.302  & 1.224  & 1.216  & 1.196  & 1.176 \\
\enddata
\end{deluxetable}

\clearpage

\begin{deluxetable}{ccccrrcrccrccrrll}
\tablecolumns{17}
\tablenum{3}
\tablewidth{0in}
\tabletypesize{\scriptsize}
\rotate
\tablecaption{Catalog of HK-Survey Field Horizontal-Branch Candidates}
\tablehead{
\colhead{Star} & \colhead{BC} & \multicolumn{2}{c}{RA (2000) DEC} & \colhead{$l$} &\colhead{$b$}& \colhead{$V$} & \colhead{$B-V$} & \colhead{EBVA}  & \colhead{$J$} & \colhead{$J-H$} & \colhead{$V_0$}  &  \colhead{$V_{0,2M}$}  &  \colhead{$BV_0$} &  \colhead{$BV_{0,2M}$} & \colhead{PC} & \colhead{CLASS}
}
\startdata
BS~15620-008  & 5  &14:50:14.2  &$+$46:21:11 &  80.2 & $+$59.7& \nodata & \nodata &  0.016 &   13.955 &    0.107 & \nodata &  14.342 & \nodata &    0.135 & H &         \\  
BS~15620-011  & 4  &14:49:52.4  &$+$44:43:14 &  77.4 & $+$60.5& \nodata & \nodata &  0.026 &   13.909 & $-$0.058 & \nodata &  13.838 & \nodata & $-$0.030 & H &         \\
BS~15620-013  & 6  &14:52:09.5  &$+$43:23:33 &  74.6 & $+$60.7& \nodata & \nodata &  0.023 &   12.777 &    0.391 & \nodata &  14.067 & \nodata &    0.681 & L &         \\
BS~15620-015  & 4  &14:50:16.6  &$+$43:01:11 &  74.2 & $+$61.2& \nodata & \nodata &  0.018 &   12.523 &    0.146 & \nodata &  13.066 & \nodata &    0.188 & H &         \\
BS~15620-017  & 5  &14:52:15.0  &$+$43:01:38 &  73.9 & $+$60.9& \nodata & \nodata &  0.022 &   15.141 &    0.039 & \nodata &  15.289 & \nodata &    0.051 & H &         \\
BS~15620-018  & 4  &14:54:38.5  &$+$43:54:57 &  75.2 & $+$60.1& \nodata & \nodata &  0.023 &   13.145 &    0.105 & \nodata &  13.539 & \nodata &    0.128 & H &         \\
BS~15620-019  & 4  &14:52:40.3  &$+$45:39:40 &  78.6 & $+$59.7& \nodata & \nodata &  0.017 &   11.608 &    0.142 & \nodata &  12.174 & \nodata &    0.183 & H &         \\
BS~15620-020  & 4  &14:53:13.9  &$+$47:02:40 &  80.9 & $+$58.9&   13.16 &    0.39 &  0.019 &   12.158 &    0.145 &  13.101 &  12.710 &   0.371 &    0.187 & M &         \\
BS~15620-021  & 5  &14:55:37.3  &$+$44:18:32 &  75.8 & $+$59.8& \nodata & \nodata &  0.026 &   14.723 &    0.023 & \nodata &  14.834 & \nodata &    0.033 & H &         \\
BS~15620-023  & 4  &14:55:58.1  &$+$43:30:27 &  74.2 & $+$60.0& \nodata & \nodata &  0.018 &   12.155 &    0.184 & \nodata &  12.822 & \nodata &    0.249 & M &         \\
BS~15620-024  & 2  &14:57:12.1  &$+$43:08:10 &  73.4 & $+$60.0& \nodata & \nodata &  0.016 &   10.775 &    0.106 & \nodata &  11.268 & \nodata &    0.134 & H &         \\
BS~15620-025  & 3  &14:56:41.2  &$+$43:04:11 &  73.3 & $+$60.1& \nodata & \nodata &  0.016 &   12.369 & $-$0.003 & \nodata &  12.510 & \nodata &    0.013 & H &         \\
BS~15620-032  & 4  &15:03:42.0  &$+$43:57:33 &  74.0 & $+$58.6& \nodata & \nodata &  0.018 &   12.473 &    0.184 & \nodata &  13.129 & \nodata &    0.249 & M &         \\
BS~15620-034  & 3  &15:05:03.3  &$+$44:32:15 &  74.9 & $+$58.1& \nodata & \nodata &  0.019 &   11.072 &    0.294 & \nodata &  12.113 & \nodata &    0.458 & L &         \\
BS~15620-035  & 6  &15:05:16.5  &$+$44:52:31 &  75.5 & $+$58.0& \nodata & \nodata &  0.018 &   13.139 &    0.359 & \nodata &  14.322 & \nodata &    0.606 & L &         \\
BS~15620-038  & 5  &15:06:24.7  &$+$45:56:51 &  77.2 & $+$57.4& \nodata & \nodata &  0.019 &   14.092 &    0.122 & \nodata &  14.512 & \nodata &    0.154 & H &         \\
BS~15620-040  & 5  &15:07:56.2  &$+$43:57:16 &  73.5 & $+$57.9& \nodata & \nodata &  0.018 &   13.656 &    0.136 & \nodata &  14.132 & \nodata &    0.175 & H &         \\
BS~15620-041  & 5  &15:07:43.3  &$+$43:38:18 &  73.0 & $+$58.0& \nodata & \nodata &  0.017 &   11.260 &    0.493 & \nodata &  12.969 & \nodata &    0.970 &L: &         \\
BS~15620-045  & 5  &15:11:25.3  &$+$43:35:10 &  72.5 & $+$57.4& \nodata & \nodata &  0.017 &   13.632 &    0.184 & \nodata &  14.253 & \nodata &    0.252 & M &         \\
BS~15620-049  & 6  &15:09:55.8  &$+$44:59:45 &  75.1 & $+$57.2& \nodata & \nodata &  0.022 &   13.336 &    0.337 & \nodata &  14.433 & \nodata &    0.551 & L &         \\
BS~15620-058  & 6  &14:56:13.8  &$+$43:42:50 &  74.6 & $+$59.9& \nodata & \nodata &  0.019 &   12.932 &    0.286 & \nodata &  13.885 & \nodata &    0.442 & L &         \\
BS~15620-060  & 5  &14:58:08.9  &$+$42:21:23 &  71.8 & $+$60.1& \nodata & \nodata &  0.019 &   12.626 &    0.328 & \nodata &  13.724 & \nodata &    0.533 & L &         \\
BS~15620-064  & 5  &14:58:39.3  &$+$42:59:28 &  72.9 & $+$59.8& \nodata & \nodata &  0.015 &   14.576 & $-$0.098 & \nodata &  14.401 & \nodata & $-$0.051 & H &         \\
BS~15620-065  & 5  &15:01:13.0  &$+$42:56:43 &  72.5 & $+$59.4& \nodata & \nodata &  0.018 &   12.833 &    0.248 & \nodata &  13.671 & \nodata &    0.365 & M &         \\
BS~15620-067  & 6  &15:05:37.8  &$+$44:25:27 &  74.6 & $+$58.1& \nodata & \nodata &  0.020 &   11.677 &    0.538 & \nodata &  13.528 & \nodata &    1.106 &L: &         \\
BS~15620-069  & 6  &15:07:49.6  &$+$43:04:24 &  72.0 & $+$58.2& \nodata & \nodata &  0.013 &   12.867 &    0.361 & \nodata &  14.076 & \nodata &    0.616 & L &         \\
\\
BS~15621-002  & 3  &10:29:10.9  &$+$25:58:56 & 206.8 & $+$58.3&   11.84 &    0.14 &  0.019 &   11.410 &    0.186 &  11.781 &  12.108 &   0.121 &    0.253 & H & A-V     \\
BS~15621-005  & 6  &10:29:25.5  &$+$23:04:24 & 212.2 & $+$57.7&   12.86 &    0.23 &  0.023 &   12.404 &    0.101 &  12.789 &  12.812 &   0.207 &    0.122 & M & A-V     \\
BS~15621-007  & 5  &10:27:58.1  &$+$27:48:06 & 203.3 & $+$58.3&   12.14 &    0.72 &  0.023 &   10.849 &    0.373 &  12.069 &  12.144 &   0.697 &    0.634 &L: &         \\
BS~15621-008  & 2  &10:27:38.2  &$+$27:48:06 & 203.3 & $+$58.2&   11.24 & $-$0.02 &  0.022 &   11.190 & $-$0.037 &  11.172 &  11.272 &$-$0.042 & $-$0.016 & H & FHB/A   \\
\enddata
\tablecomments{Table 3 is published in its entirety in the electronic edition.  A portion is shown here for guidance regarding its form and content.}
\end{deluxetable}

\clearpage

\begin{deluxetable}{ccccl}
\tablecolumns{5}
\tablenum{4}
\tablewidth{0pt}
\tabletypesize{\scriptsize}
\tablecaption{2MASS Identifications  of HK-Survey Field Horizontal-Branch Candidates and Multiple Identifications}
\tablehead{
\colhead{Star} & \colhead{2MASS ID} & \multicolumn{2}{c}{RA (2000) DEC} &\colhead{MUTIPLE ID} 
}
\startdata
BS~15620-008 & 14501408+4621106 & 14:50:14.09 & $+$46:21:10.7 &                                                  \\                           
BS~15620-011 & 14495209+4443132 & 14:49:52.09 & $+$44:43:13.2 &                                                  \\                                 
BS~15620-013 & 14520936+4323336 & 14:52:09.37 & $+$43:23:33.6 &                                                  \\                                 
BS~15620-015 & 14501637+4301107 & 14:50:16.38 & $+$43:01:10.7 &                                                  \\                                 
BS~15620-017 & 14521480+4301411 & 14:52:14.80 & $+$43:01:41.2 &                                                  \\                                 
BS~15620-018 & 14543833+4354584 & 14:54:38.33 & $+$43:54:58.5 &                                                  \\                                 
BS~15620-019 & 14524022+4539408 & 14:52:40.22 & $+$45:39:40.9 &                                                  \\                                 
BS~15620-020 & 14531379+4702434 & 14:53:13.79 & $+$47:02:43.4 &    BS~16083-043                                  \\
BS~15620-021 & 14553720+4418320 & 14:55:37.21 & $+$44:18:32.1 &                                                  \\
BS~15620-023 & 14555801+4330281 & 14:55:58.01 & $+$43:30:28.1 &                                                  \\
BS~15620-024 & 14571204+4308107 & 14:57:12.05 & $+$43:08:10.7 &                                                  \\
BS~15620-025 & 14564123+4304097 & 14:56:41.23 & $+$43:04:09.7 &                                                  \\
BS~15620-032 & 15034183+4357336 & 15:03:41.84 & $+$43:57:33.6 &                                                  \\
BS~15620-034 & 15050328+4432136 & 15:05:03.28 & $+$44:32:13.7 &                                                  \\
BS~15620-035 & 15051651+4452310 & 15:05:16.52 & $+$44:52:31.0 &                                                  \\
BS~15620-038 & 15062450+4556501 & 15:06:24.50 & $+$45:56:50.2 &                                                  \\
BS~15620-040 & 15075627+4357165 & 15:07:56.27 & $+$43:57:16.5 &                                                  \\
BS~15620-041 & 15074332+4338161 & 15:07:43.32 & $+$43:38:16.1 &                                                  \\
BS~15620-045 & 15112521+4335079 & 15:11:25.22 & $+$43:35:07.9 &                                                  \\
BS~15620-049 & 15095560+4459430 & 15:09:55.60 & $+$44:59:43.1 &                                                  \\
BS~15620-058 & 14561368+4342504 & 14:56:13.69 & $+$43:42:50.5 &                                                  \\
BS~15620-060 & 14580896+4221231 & 14:58:08.96 & $+$42:21:23.1 &                                                  \\
BS~15620-064 & 14583921+4259277 & 14:58:39.22 & $+$42:59:27.7 &                                                  \\
BS~15620-065 & 15011296+4256417 & 15:01:12.97 & $+$42:56:41.8 &                                                  \\
BS~15620-067 & 15053776+4425271 & 15:05:37.77 & $+$44:25:27.2 &                                                  \\
BS~15620-069 & 15074962+4304220 & 15:07:49.62 & $+$43:04:22.0 &                                                  \\
\\
BS~15621-002 & 10291095+2558531 & 10:29:10.96 & $+$25:58:53.1 &                                                  \\
BS~15621-005 & 10292557+2304191 & 10:29:25.57 & $+$23:04:19.2 &                                                  \\
BS~15621-007 & 10275807+2748054 & 10:27:58.07 & $+$27:48:05.4 &                                                  \\
BS~15621-008 & 10273817+2748044 & 10:27:38.17 & $+$27:48:04.4 &                                                  \\
\enddata
\tablecomments{Table 4 is published in its entirety in the electronic edition.  A portion is shown here for guidance regarding its form and content.}
\end{deluxetable}

\clearpage

\begin{deluxetable}{cccc}
\tablecolumns{4}
\tablewidth{0pt}
\tablenum{5}
\tablecaption{Comparison with the Wilhelm et al. (1999b) Classifications}
\tablehead
{
\colhead{Likelihood Class} & \colhead{Number} & \colhead{Class FHB or FHB/A} & Percentage
}
\startdata
 H    &     388    &   303   &  78\% \\
 M    &     240    &    90   &  37\% \\
 L    &      33    &    10   &  30\% \\
\enddata
\end{deluxetable}


\begin{thebibliography}{}

\bibitem[Adelman-McCarthy et al.(2006)]{2006ApJS..162...38A}
Adelman-McCarthy, J.~K., et al. 2006, \apjs, 162, 38

\bibitem[Arnold \& Gilmore(1992)]{1992MNRAS.257..225A} Arnold, R., \&
Gilmore, G. 1992, \mnras, 257, 225

\bibitem[Beers, Preston, \& Shectman(1988)]{1988ApJS...67..461B} Beers, T.~C.,
Preston, G.~W., \& Shectman, S.~A. 1988, \apjs, 67, 461 (FHB~I)

\bibitem[]{} Beers, T.C., Flynn, K., and Gebhardt, K. 1990, \aj, 100, 32

\bibitem[Beers et al. (1996)]{1996ApJS..103..433B} Beers, T.~C., Wilhelm, R.,
Doinidis, S.~P., Mattson, C.~J. 1996, \apjs, 103, 433 (FHB~II)

\bibitem[]{} Beers, T.C., et al. 2006, \apjs, in press (astro-ph/0610018)

\bibitem[Bonifacio, Monai,\& Beers (2000)]{ 2000AJ....120.2065B}Bonifacio, P., Monai, S., \& Beers, T.~C. 2000, \aj, 120, 2065

\bibitem[Brown et al. (2004)]{2004AJ....127.1555B} Brown, W.~R., Geller, M.~J., Kenyon, S.~J., Beers, T.~C., Kurtz, M.~J., \& Roll, J.~B. 2004, \aj, 127, 1555

\bibitem[Christlieb, N.(2003)]{2003RvMA...16..191C} Christlieb, N. 2003, Rev. Mod. Astron., 16, 191

\bibitem[Christlieb et al.(2005)]{ 2005A&A...431..143C} Christlieb, N., Beers,
 T.~C., Thom, C., Wilhelm, R., Rossi, S., Flynn, C., Wisotzki, L., Reimers, D.  2005, A \& A, 431, 143

\bibitem[Clewley et al.(2004)]{2004MNRAS.352..285C} Clewley, L., Warren,
S.~J., Hewett, P.~C., Norris, J.~E., \& Evans, N.~W. 2004, \mnras,
352, 285

\bibitem[Doinidis \& Beers(1989)]{1989ApJ...340L..57D} Doinidis, S.~P., \&
Beers, T.~C. 1989, \apjl, 340, L57

\bibitem[Doinidis \& Beers(1990)]{ 1990PASP..102.1392D} Doinidis, S.~P., \&
Beers, T.~C. 1990, \pasp, 102, 1392

\bibitem[Doinidis \& Beers(1991)]{1991PASP..103..973D} Doinidis, S.~P., \&
Beers, T.~C. 1991, \pasp, 103, 973

\bibitem[Flynn et al.(1995)]{1995A&AS..109..171F} Flynn, C., Sommer-Larsen,
J., Christensen, P.~R., \& Hawkins, M.~R.~S. 1995, \aaps, 109, 171

\bibitem[]{} Girard, T.M., Dinescu, D.I., van Altena, W.F., Platais, I.,
Monet, D.,\& Lopez, C.E. 2004, \aj, 127, 3060

\bibitem[Kinman et al.(1994)]{1994AJ....108.1722K} Kinman, T.~D., Suntzeff,
N.~B., \& Kraft, R.~P. 1994, \aj, 108, 1722

\bibitem[MacConnell, Sthepheson, \& Pesch (1993)]{1993ApJS...86..453M} MacConnell, D.~J., Stephenson, C.~B., Pesch, P. 1993, \apjs, 86, 453

\bibitem[Martin et al.(2005)]{2005ApJ...619L...1M} Martin, D.~C., et al.
2005, \apjl, 619, L1

\bibitem[]{} Munn, J.A. et al. 2004, \aj, 127, 3034

\bibitem[Newberg et al. (2003)]{2003ApJ...596L.191N} Newberg, H.~J. et al.  2003, \apj, 596, L191

\bibitem[Norris \& Hawkins(1991)]{1991ApJ...380..104N} Norris, J.~E., \&
Hawkins, M.~R.~S. 1991, \apj, 380, 104

\bibitem[Norris, Ryan \& Beers (1999)]{ 1999ApJS..123..639N}Norris, J.~E., Ryan, S.~G., \& Beers, T.~C.  1999, \apjs, 123,639

\bibitem[Pier(1982)]{1982AJ.....87.1515P} Pier, J.~R. 1982, \aj, 87, 1515

\bibitem[]{} Pier, J.R. 1984, \apj, 281, 260

\bibitem[Preston, Shectman \& Beers (1991)]{1991ApJ...375..121P} Preston, G.~W., Shectman, S.~A., \& Beers, T.~C. 1991, \apj, 375, 121

\bibitem[Rodgers et al.(1993)]{1993AJ....106..591R} Rodgers, A.~W.,
Roberts, W.~H., \& Walker, I. 1993, \aj, 106, 591

\bibitem[Sakamoto et al. (2003)]{2003A&A...397..899S} Sakamoto, T., Chiba, M., \& Beers, T.~C.  2003, A\&A, 397, 899

\bibitem[Schlegel, Finkbeiner, \& Davis(1998)]{1998ApJ...500..525S} Schlegel, D.~J., Finkbeiner, D.~P., \& Davis, M.  1998, \apj, 500, 525

\bibitem[Sirko et al.(2004a)]{2004AJ....127..899S} Sirko, E., et al.  2004a, \aj, 127, 899

\bibitem[Sirko et al.(2004b)]{2004AJ....127..914S} Sirko, E., et al. 2004b,
\aj, 127, 914

\bibitem[Skrutskie et al.(2006)]{2006AJ....131.1163S} Skrutskie, M.~F., et
al. 2006, \aj, 131, 1163

\bibitem[Sommer-Larsen et al. (1997)]{1997ApJ...481..775S} Sommer-Larsen, J., Beers, T.~C., Flynn, C., Wilhelm, R., \& Christensen, P. R.  1997, \apj, 481, 775

\bibitem[Yanny et al.(2000)]{2000ApJ...540..825Y} Yanny, B., et al. 2000,
\apj, 540, 825

\bibitem[York et al.(2000)]{2000AJ....120.1579Y} York, D.~G., et al. 2000,
\aj, 120, 1579

\bibitem[Thom et al. (2005)]{2005MNRAS.360..354T} Thom, C., Flynn, C., Bessell, M.~S., H\"anninen, J., Beers, T.~C.; Christlieb, N., James, D., Holmberg, J., \& Gibson, B.~K.  2005, \mnras, 360, 354

\bibitem[Thom et al. (2006)]{ 2006ApJ...638L..97T} Thom, C., Putman, M.~E.,
Gibson, B.~K., Christlieb, N., Flynn, C., Beers, T.~C., Wilhelm, R., \&  Lee,
 Y.~S. 2006, \apj, 638, L97 

\bibitem[]{} Wakker, B. P., \& van Woerden, H. 1997, ARA\&A, 35, 217 

\bibitem[Wilhelm et al. (1999a)]{ 1999AJ....117.2308W} Wilhelm, R., Beers,
 T.~C., \& Gray, R. O. 1999a, \aj, 117, 2308

\bibitem[Wilhelm et al. (1999b)]{ 1999AJ....117.2329W} Wilhelm, R., Beers,
 T.~C., Sommer-Larsen, J., Pier, J.~R., Layden, A.~C., Flynn, C., Rossi, S., \& 
Christensen, P.~R. 1999, \aj, 117, 2329

\bibitem[]{} Zacharias, N., Urban, S. E., Zacharias, M. I., Wycoff, G. L., Hall, D. M.,
Monet, D. G., \& Rafferty, T. J. 2004, \aj, 127, 3043

\end{thebibliography}
\end {document}